\documentclass[3p,times]{elsarticle}
\usepackage{ecrc_dummy}
\volume{00}
\firstpage{1}
\runauth{}
\usepackage{amssymb}

\usepackage[figuresright]{rotating}
\usepackage{amsmath}
\begin{document}

\begin{frontmatter}

\title{A ring lasers array for fundamental physics}

\author{
A. Di Virgilio$^{a}$,
M. Allegrini$^{b}$,
A. Beghi$^{c}$
J. Belfi$^{a}$,            
N. Beverini$^{b}$,         
F. Bosi$^{a}$,  
B. Bouhadef$^{a}$,
M. Calamai$^{b}$,
G. Carelli$^{b}$,
D. Cuccato$^{c, d}$,  E. Maccioni$^{b}$, A. Ortolan$^{d}$, G. Passeggio$^{e}$, A. Porzio$^{e, f}$, M.L. Ruggiero$^{g}$, R. Santagata$^{h}$ and A. Tartaglia$^{g}$}          

\address[$^a$] {INFN Sezione di Pisa, Pisa Italy}
\address[$^b$] { Department of Physics, University of Pisa, Pisa Italy}
\address[$^c$] {DEI, University of Padua, Padua Italy}
\address[$^d$] {INFN-Legnaro National Laboratory, Legnaro (Padua) Italy}
\address[$^e$] {INFN Sezione di Napoli, Naples Italy}
\address[$^f$] {CNR-SPIN, Naples Italy}
\address[$^g$] {Polythecnic of Turin, Turin Italy}
\address[$^h$] {Department of Physic, University of Siena, Siena Italy}

\date{ }

\begin{abstract}
After reviewing the importance of light as a probe for testing the structure of space-time, we describe the GINGER project. GINGER will be a three-dimensional array of large size ring-lasers able to measure the de Sitter and Lense-Thirring effects. The instrument will be located at the underground laboratory of GranSasso, in Italy. We describe the preliminary actions and measurements already under way and present the full road map to GINGER. The intermediate apparatuses GP2 and GINGERino are described. GINGER is expected to be fully  operating in few years.
\end{abstract}

\end{frontmatter}

\section{Introduction}
\label{sec-1}

One of the pillars of contemporary understanding of matter, energy and
space-time is general relativity (GR). Its successes in explaining the
behaviour of the world around us and of the whole universe are well known as
well as its so far unresolved conflict with quantum mechanics in the high
energy domain. It is however true that also in the very low energy sector of
the gravitational interaction there are predictions of GR which have not
been fully explored up to these days.

A typical example is the so called gravito-magnetic component of the
gravitational field, whose direct verification relies for the moment on
three only experiments in space: Gravity Probe B (GP-B) that took data from
2004 to 2005 and was concluded and the results published in 2011\cite{GP-B};
the two LAGEOS satellites orbital nodes analysis, published in 2004\cite%
{LAGEOS1} and, with an improved modeling of the gravitational field of the
Earth, in 2011\cite{LAGEOS2}; the LARES\ mission, under way and gathering
data, launched in February 2012 \cite{LARES}.

GP-B verified the geodetic effect in the gravitational field of the Earth
with an accuracy of 0.28\% and the Lense-Thirring (LT) drag with an accuracy
of 19\%; the analysis of the precession of the nodes of the LAGEOS
satellites verified the LT effect with the accuracy of 10\%; finally LARES
is working to determine the LT drag with an accuracy of a few \% (possibly
1\%). Other evidence of gravito-magnetic effects may be found in the laser
ranging of the orbit of the moon and in the study of the dynamics of binary
systems composed of at least one compact massive object (neutron star).

Another example of a weak effect predicted by GR are gravitational waves. No
direct measurement has been performed so far, but very strong indirect
evidence for their existence is obtained from the observation of double star
systems including a pulsar \cite{perioddec}.

Besides pure GR\ effects the observation of the universe on the widest scale
provides also facts which can be consistent with GR assuming that otherwise
unseen entities exist, such as dark matter and dark energy. The former would produce the additional gravity
required to explain the rotation curves of galaxies and the speeds of the
components of star or galaxy clusters. The latter would be necessary to
generate the push required by the accelerated expansion of the universe.
These facts, partly going back to the thirties of the last century (dark
matter) \cite{dark matter}, partly quite recent (dark energy) \cite{dark
energy}, have stimulated ideas implying that GR might need some extension if
not a complete change of paradigm. What matters here is that the
phenomenology to look for and to analyze in search for differences from GR
is in the domain of low and ultralow energies.
The above remarks present reasonable motivations for working experimentally
on the gravitational interaction in the weak domain looking for
post-Newtonian effects and Parametrized Post Newtonian (PPN) descriptions
which could evidence deviations from classical GR. Can such an investigation
be conducted in a laboratory, besides relying on large scale observation of
the sky? The answer is yes and, among various possible experimental
approaches, a perfect tool is represented by light. Light is indeed
intrinsically relativistic and is affected in various ways by the
gravitational field. In the classical domain and, as far as a theory is
considered treating space-time as a continuous four-dimensional Riemannian
manifold, light completely covers the manifold with a network of null
geodesics. If we find the way of reading the local and global configuration
of the null geodesics tissue we can reconstruct the "shape" of space-time
i.e. the gravitational field and see whether it fully corresponds to the GR
description or maybe there is something missing.

While considering how to exploit light in order to explore the gravitational
field we should add that the advancement of the laser technologies has
pushed the possibilities of such devices to unprecedented values of accuracy
and precision. All in all, a laser, and in particular a ring-laser, appears
today as a most interesting apparatus to probe the structure of space-time
at the laboratory scale.

These are the main motivations for the design and implementation of an
experiment based on the use of ring lasers for fundamental physics. The main
purpose is to explore the asymmetric propagation of light along a closed
space path in the gravitational field of a rotating body. In a sense, the
prototype of this type of experiments is the old Sagnac interferential
measurement of what we can now call the \textit{kinematic }asymmetry of the
propagation of light along a closed space contour as seen by a rotating
observer in a flat space-time (no gravitational field) \cite{Sagnac}. So far
ring lasers have been built as Sagnac sensors of absolute rotations (which
means with respect to the "fixed stars") for practical purposes, as compact
and sensible devices replacing mechanical gyroscopes (this is the reason why
ring lasers are also called gyrolasers) for navigation or, in the case of
the most refined instruments, for geodesy or even for determining the Length
Of the Day in competition with VLBI (Very Long Base Interferometry). The
latter application, which is fundamental, is already in the reach of the G
Ring in Wettzell \cite{G Ring}. The latter facility is by now on the verge
of being able to detect not only the kinematical rotations of the
laboratory, but also the physical effects of the gravitational field due to
the rotation of the source and of the laboratory. 

Our experiment, named GINGER (Gyroscopes IN GEneral Relativity), is intended
to further improve the technology beyond G. The rest of the paper will
describe both the theoretical framework and the final configuration of
GINGER, and the steps which are under way in order to test the innovative
technologies we are going to use and in order to build the final laboratory
which will be located in the LNGS (Gran Sasso National Laboratories) of the
Italian INFN. The ring-laser appears today as a most interesting apparatus
to probe the structure of space-time at the laboratory scale. At his early
stage, the expected sensitivity of GINGER will not be competitive with space
measurements to test PPN theories, but being the apparatus on Earth,
improvements will be feasible with time.

\section{Light in the gravitational field of a rotating body}
\label{sec-2}

Assuming that space-time can be described by a metric theory on a
four-dimensional Riemannian manifold with Lorentzian signature, the central
geometric object containing the essence of the gravitational field will be
the line element. If the source of curvature (i.e. gravity) is a steadily
and freely rotating object the line element is given by:

\begin{equation}
ds^{2}=g_{00}c^{2}dt^{2}+g_{rr}dr^{2}+g_{\theta \theta }r^{2}d\theta
^{2}+g_{\phi \phi }r^{2}\sin ^{2}\theta d\phi ^{2}+2g_{0\phi }cr\sin \theta
d\phi dt  \label{line element}\ .
\end{equation}

The coordinates used in (\ref{line element}) are polar in space with the
radial coordinate measured from the barycenter of the central mass, assumed
to be in free fall, the $\theta $ angle (colatitude) measured from the
rotation axis of the source and $\phi $ (longitude) measured from a fixed
direction (with respect to the "fixed stars") in space; time $t$ is measured
by clocks located in a remote region not influenced by the gravitational
field. The not fully standard notation used in (\ref{line element}) insures
the dimensionlessness of the $g_{\mu \nu }$ functions; the speed of light $c$
is here essentially a conversion factor transforming time into a length. The
$g_{\mu \nu }$'s (the components of the metric tensor) depend on the
variables $r$ and $\theta $ only, because of the symmetry. If the central
mass is indeed rotating no global coordinate transformation exists
converting the metric in (\ref{line element}) to the Minkowski metric.

The underlying assumptions so far are:

\begin{itemize}
\item the source of gravity is isolated;

\item the central object is rigid or at least it keeps its shape and mass
distribution fixed in time;

\item space-time is asymptotically flat and Minkowskian.\newline
\end{itemize}

If we consider a real system, such as the terrestrial gravitational field,
none of the above conditions, strictly speaking, is satisfied. The earth is
influenced by the other bodies in the solar system so that its axis does not
keep a fixed orientation with respect to the quasars ("fixed stars"). The
gravitational perturbations induced by the surrounding bodies and the
differential heating of the surface cause changes in the shape and mass
distribution because of the non-rigidity of the planet. Space-time is not
flat anywhere in the universe because no empty asymptotic region exists.

If we are interested in tiny relativistic effects we shall be very careful
while using the simple symmetries implied in (\ref{line element}), because
they are all imperfect.

In any case, working with light and assuming that $c$ is the same for all
freely falling observers (which is the essence of relativity), the
corresponding line element will be equal to $0$ and we will be able to write
the coordinated time span along the world line of a light ray as:

\begin{equation}
dt=\frac{-g_{0\phi }r\sin \theta d\phi \pm \sqrt{g_{0\phi }^{2}r^{2}\sin
^{2}\theta d\phi ^{2}-g_{00}\left( g_{rr}dr^{2}+g_{\theta \theta
}r^{2}d\theta ^{2}+g_{\phi \phi }r^{2}\sin ^{2}\theta d\phi ^{2}\right) }}{%
cg_{00}}\ .
  \label{tempo}
\end{equation}

To ensure in any case an evolution towards the future ($dt>0$) the $+$ sign
must be chosen  when going toward increasing $\phi$'s and the - sign when moving in the opposite direction.

Equation (\ref{tempo}) permits to evaluate the coordinated time of flight of
an electromagnetic signal between two successive events in vacuo. Let us
consider a closed path (in space); of course, excluding the horizon of a
black hole, a closed path may be followed by light only in presence of some
technical expedient (mirrors, optical fiber).

Integrating over the path, both on the right and on the left, two different
results are obtained because of the off diagonal $g_{0\phi }$ component of
the metric tensor. Let us use the angular velocity of the central body as a
reference for the rotation sense: the so defined anticlockwise sense will
correspond to $d\phi >0,$ the clockwise will correspond to $d\phi <0$.
Finally we see that the difference between the corotating time of flight, $%
t_{+}$, and the counter-rotating one, $t_{-}$, will be:

\begin{equation}
\delta t=t_{+}-t_{-}=-\frac{2}{c} \oint \frac{g_{0\phi }}{g_{00}}r\sin
\theta d\phi\ .
  \label{diff coord}
\end{equation}

If at the start and arrival point, imagined as being fixed in the chosen
reference frame, there is a device sensible to (\ref{diff coord}) its proper
time $\tau $ difference will be:

\begin{equation}
\delta \tau =-\frac{2}{c}\sqrt{g_{00}} \oint \frac{g_{0\phi }}{g_{00}}r\sin
\theta d\phi\ .
  \label{proper diff}
\end{equation}

The $\delta \tau $ difference is the basis of the way a ring laser works. $%
\delta \tau $ may be measured letting the two counter-rotating beams
interfere and this is the typical Sagnac technique; the way of a ring laser
is however different. Since the emission of light is continuous and steady,
two standing waves, associated with the two rotation senses, are formed and
co-exist in the annular cavity of the laser. The time of flight difference
is converted into different frequencies of the two waves, and in turn the
frequency difference gives rise to a beat note. The frequency of the beat
can be read analysing the power spectrum of the signal extracted at any
point of the ring. The beat frequency $f _{b}$ is

\begin{equation}
f_{b}=c^{2}\frac{\delta \tau }{2P\lambda }=-\frac{c}{P\lambda }\sqrt{g_{00}}%
\oint \frac{g_{0\phi }}{g_{00}}r\sin \theta d\phi  \label{beat}\ ,
\end{equation}

where $P$ is the length of the perimeter of the ring and $\lambda $ is the
wavelength of the radiation.

\subsection{A laboratory on Earth}
\label{sub-1}

So far we have assumed an observer at rest with respect to the fixed stars,
which is a quite unphysical situation. In practice the experiment we want to
perform will be located within a laboratory fixed to the solid body of the
Earth. If so, we should update our choice of the coordinate system. There
are various possibilities; the simplest probably is to still choose a global
reference frame, but let its axes rotate together with the Earth. In this
way basically the coordinates remain the same but colatitude and longitude
are terrestrial rather than celestial.

Now from the viewpoint of the fixed stars the paths followed by the light
beams we want to use are no longer closed in space, because of the motion of
the laboratory, but they still turn out to be closed in the corotating
terrestrial reference frame. The general form of the line element still is
like (\ref{line element}), but now the functions have different forms. Under
the same assumptions as before we may work out the new metric elements
applying first a kinematical rotation of the axes at the angular speed of
the Earth $\Omega $, then a physical Lorentz boost at the peripheral speed
of the Earth in correspondence of the location of the laboratory \cite{noiPR}.

The formal result of these two steps is a bit complicated, but for practical
purposes we may approximate the result considering that:

\begin{eqnarray}
\frac{\Omega R}{c} &\sim &10^{-6}  \notag \\
G\frac{M}{c^{2}R} &=&\frac{\mu }{R}\sim 10^{-9}  \label{parametri} \\
G\frac{J}{c^{3}R^{2}} &=&\frac{j}{R^{2}}\sim 10^{-15}  \notag
\end{eqnarray}

$G$ is Newton's constant; $M$ is the mass of the Earth; $R$ is its radius at
the location of the laboratory and $J$ is the angular momentum of our planet.

The highest order to which we are interested is the lowest non-zero term
containing $j/R^{2}$. Under this condition the final beat frequency $f_{b}$
turns out to be:

\begin{equation}
f_{b}\simeq 2\frac{A}{\lambda P}\Omega \left( \hat{u}_{a}\cdot \hat{u}%
_{n}\right) +\frac{cA}{\lambda PR}\left( 2\left( \frac{\Omega \mu }{c}\sin
\theta -\frac{j}{R^{2}}\cos \theta \right) \left( \hat{u}_{r}\cdot \hat{u}%
_{n}\right) -\frac{j}{R^{2}}\left( \hat{u}_{\theta }\cdot \hat{u}_{n}\right)
\sin \theta \right)\ ,
  \label{beatf}
\end{equation}

where $A$ is the area of the ring; the $\hat{u}$'s are unit vectors in the
directions, respectively, of the axis of the Earth ($_{a}$), the normal to
the ring ($_{n}$), the direction of the local meridian ($_{\theta }$). The
ratio $\frac{A}{\lambda P}$ is called the \textit{scale factor }$S$ of the
instrument. The second term on the right of the formula is approximately $%
10^{-9}$ times the first.

\section{The GINGER Project: Gyrolasers for fundamental relativity}
\label{sec-3}

Considering the orders of magnitude (\ref{parametri}) and formula (\ref%
{beatf}), we see that, in order to reveal general relativistic effects
depending on the mass and the angular momentum of the Earth, we need a
device endowed with a sensitivity at least nine orders of magnitude better
than the one required for measuring the plane angular velocity of the Earth,
through the classical Sagnac effect. In fact in formula (\ref{beatf}) the
first term is the classical Sagnac term, whereas the second contains both
the Lense-Thirring drag, depending on the angular momentum $\bar{j}$, and
the de Sitter or geodetic term expressing the interaction of the local
Newtonian force with the angular velocity of the Earth. The latter two
contributions turn out to have the same order of magnitude on the surface of
the planet.

Is the needed sensitivity available or attainable today? Commonly, in
navigation applications, ring lasers are based on single longitudinal mode
He-Ne lasers operating at a wavelength of $632.8$~nm. Inertial navigation
devices usually have an area $<0.02$ m$^{2}$ corresponding to a perimeter of
$30$ cm or less. The typical sensitivity of such devices is around $5\times
10^{-7}$ rad$/$s$/\sqrt{\text{Hz}}$ and the drift is as low as $0.0001$%
deg/h. This performance level is fully sufficient for navigational demands
but falls short by several orders of magnitude for most geophysical
applications; the more for fundamental physics.

The Gross Ring (G) in Wettzell is a square ring, 4 m in side, mounted on an
extremely rigid and thermally stable monolithic zerodur slab, located under
an artificial 35 m thick mound. The most recent performance of G, expressed
in terms of measured equivalent angular velocity, has a lower boundary below
$1$ prad/s (picoradian/second) at 1000 s integration time \cite{ulli}. This
sensitivity is above the requirement for the measurement of the GR effects,
but various improvements in technologies, global design and signal cleaning
should fill the remaining gap. \newline
Actually at the level of prad/s and less, many delicate problems arise,
besides the ones already mentioned, concerning the stability and behaviour
of the laser and the mirrors. Formula (\ref{beatf}) has been obtained under
the hypothesis that the rotational speed of the Earth is a constant, but
this is not the case because of the coupling of the moment of inertia of the
planet with the gravitational influence of the moon and the sun, which in
turn change with the configuration of the two celestial bodies. Furthermore
the non rigidity of the Earth appears, influencing the instantaneous moment
of inertia of the planet. Even the angles appearing in (\ref{beatf}) are not
stable at the required accuracy of nrad or less, because of the non rigidity
of the crust of the Earth and because of mechanical and thermal
instabilities of the measuring device.

A specific difficulty that has to be faced in the LT measurement is that the
sought for effect is a tiny time independent quantity superposed to a
comparatively huge signal (the kinematical Sagnac term), so that the
calibration is quite demanding. For this reason an accurate investigation of
the systematics of the laser is needed, and different techniques for
extracting the signal need be considered and evaluated. The result could in
principle be validated repeating the measurement with different techniques
and operating the laser in a different working point. In fact one could also
use a passive cavity, i.e. the measurement could be repeated with the same
apparatus but using an external laser source to interrogate the array of
cavities. The technique of the passive cavity Sagnac is however not mature
as the active one, but in general a ring-laser system allows to repeat the
measurement with two different methods, having different systematics thus
making the detection of a small constant effect easier.

 G has reached remarkable sensitivity and stability, which
makes the goal of using this kind of instrumentation for fundamental physics
experiments a demanding but reasonable objective.

However, the different contributions to the beat frequency correspond to
effective rotations along different directions. In practice, in order to
discriminate the various terms, it is necessary to have a three-dimensional
device able to measure the three components of the rotation vectors. The
monolithic design cannot easily be extended to such a three-axial ring-laser
system. Not considering mechanical difficulties, the monolithic solution
would have prohibitive costs.

To overcome the above weaknesses and difficulties we have conceived the idea
of a three-dimensional array of square rings (each of which bigger than the
present G ring), mounted on a heterolitic structure. Since the control of
the shape of the rings is vital at the level of accuracy required, the
rigidity of the 'monument' carrying the mirrors and cavities would be
replaced by a dynamical control of each perimeter (like, at a smaller scale,
for G-Pisa). In practice the size and shape of any loop can be stabilized by
piezoelectric actuators applied to the holders of the mirrors. The control loop, that will drive the piezos, will also exploit the optical cavities installed along the square diagonals. In order to develop and test the above said controlled
ringlaser, GP2, a new prototype, has been realized; it is equipped with 6
piezos. The GP2 experimental set-up has been recently completed in the
laboratory of S.Piero a Grado, close to Pisa.

In addition, the final location of the laboratory could not be as close to
the surface of the Earth as in Wettzell, because of the limits imposed by
the top soil slow motion due to atmospheric pressure changes, rain, wind etc. The location could be underground at the  Gran Sasso underground laboratory
(LNGS)\ facility in Italy, in a cavern under an average rock coverage 1400 m
thick. This arrangement will insure a very good shielding against all kinds
of surface noise. 

\begin{figure}[htbp]
\centering
\includegraphics[width=8cm]{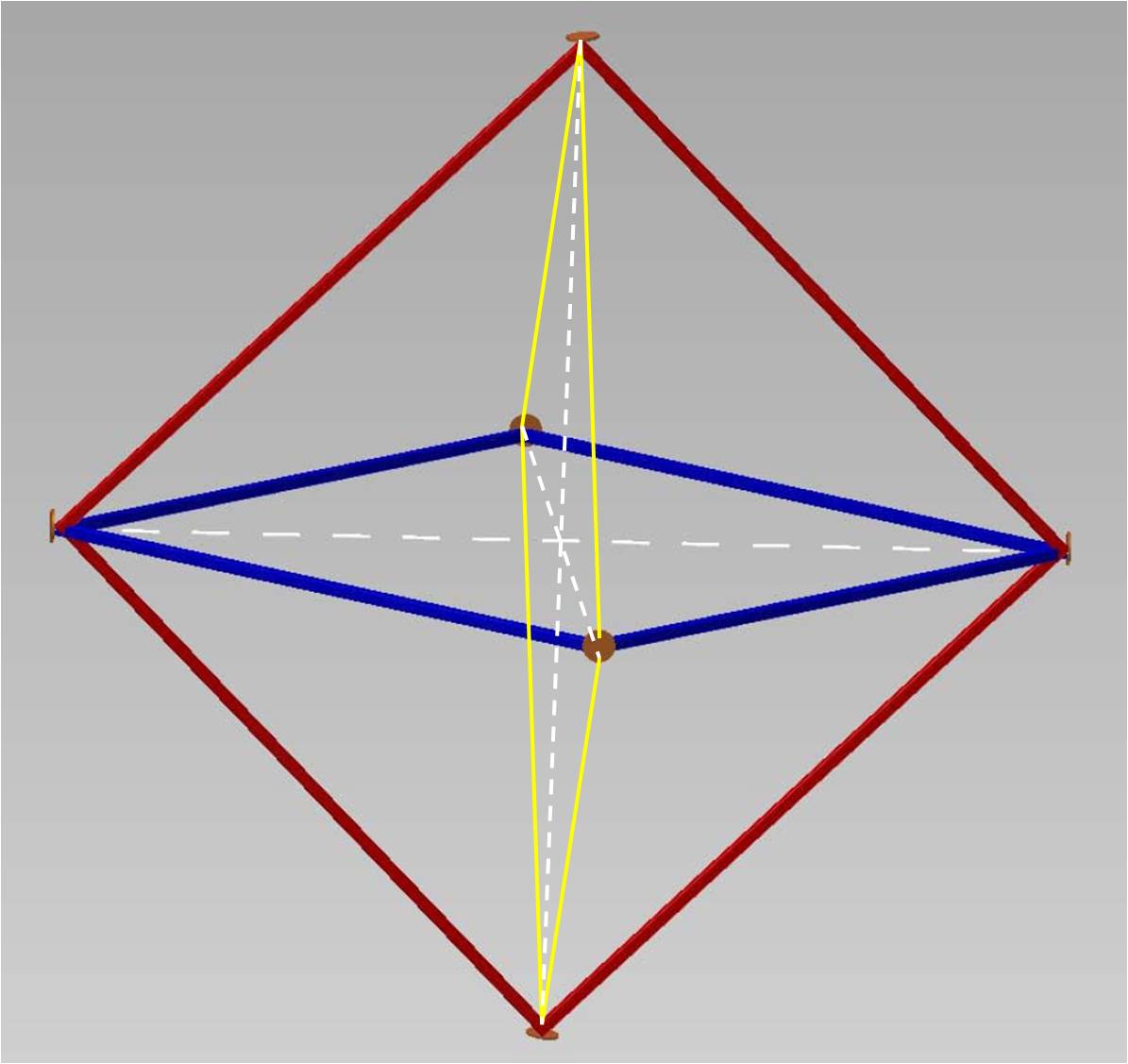}
\caption{Octahedral configuration of
GINGER. Six mirrors give rise to three mutually perpendicular square rings.
The active control of the geometry may be achieved by laser cavities along
the three diagonals connecting the mirrors.}
\label{fig-1}
\end{figure}

A possible configuration for GINGER is shown on fig. (\ref{fig-1}).
Actually the octahedral structure is the most compact, and, in principle,
easy to control, configuration; being the control obtained by means of laser
cavities along the three main diagonals of the octahedron. The side of each
of the three square loops would be not less than $6$ m.

\section{The GINGER roadmap}
\label{sec-5}

The actual building of GINGER requires a number of preliminary steps and
phases related to the technologies and measurement strategies to be
deployed. For this reason we have devised a \textit{roadmap }to GINGER.

\subsection{Goals and needs}
\label{sub-2}

\begin{figure}[htbp]
\centering
\includegraphics[width=14cm]{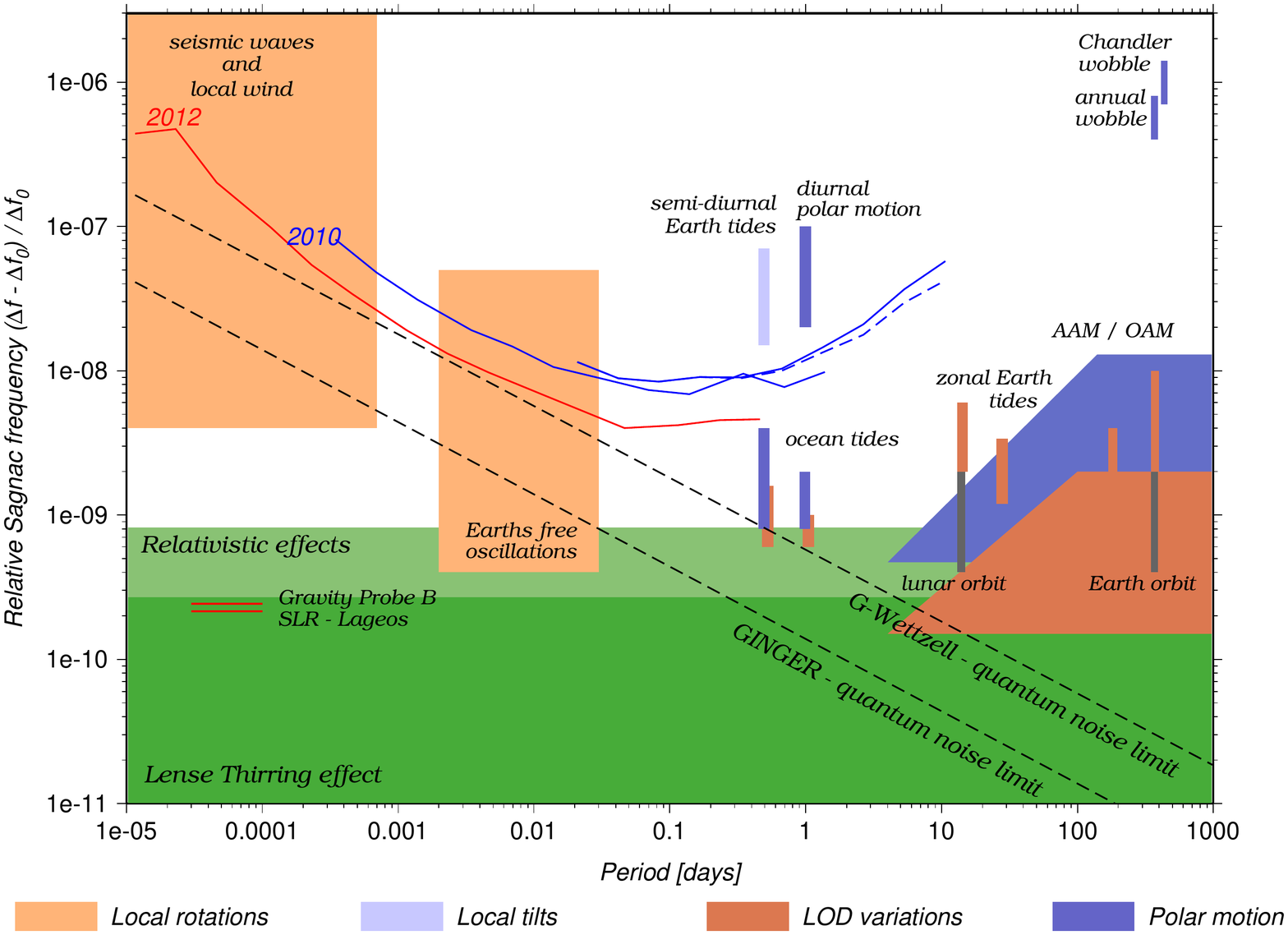}
\caption{The picture compares the Allan deviation of G in Wettzell (years 2012 and 2010, courtesy of K. U. Schreiber) with the most relevant geodetic signals, the green parts show the region of interest for the geodetic precession and the Lense-Thirring effect, on the left of the picture it is possible to see the present level of test obtained by Gravity Probe B and Lageos. The two dotted lines show the shot noise of the G (16 m perimeter), and of a ring with perimeter 24m. }
\label{allan}
\end{figure}

We assume G as a benchmark for our project. Its intrinsic structural
stability and a careful work to control the cavity length and laser
discharge parameter made it possible to obtain a stability performance very
close to the shot noise limit up to $\approx 10^{4}$~s integration time.
This corresponds to a statistical error in the angular velocity evaluation
at a level of $\mathrm{10^{-8}\times \Omega _{Earth}}$, a factor of 5 above
$\Omega _{LT}$ (the Lense-Thirring contribution to the Earth rotation), fig. \ref{allan}
 compares the results of G in Wettzel with what is necessary in order to be sensitive to the relativistic signals (for more details see K.U. Schreiber contribution to this book). 

Such
an impressive long term stability has been obtained by an accurate modelling
of all the environmental effects of geodetic, geophysical, or meteorological
origin. G lacks however absolute accuracy: it is sensitive to one component
of the angular velocity vector, and the absolute orientation of the laser
cavity with respect to the fixed stars inertial frame cannot be measured
with the required degree of precision. In order to arrive to a measurement
of the Lense-Thirring effect, we need to improve the instrumental apparatus
with respect to the following issues:

\begin{itemize}
\item[1.] The signal to noise ratio (SNR), where noise is the shot noise of the instrument, should be increased. This can be obtained by:

\begin{itemize} 
\item[a] increasing the size (the increase in SNR is more than quadratic with the size )

\item[b] improving as much as possible the quality of the mirrors, and with a careful choice of the reflectivity and transmission;

\item[c] investigating new techniques of laser operation (multimode locked
operation, split mode etc.);

\item[d] increasing as much as possibile the integration time with a suitable location of the apparatus
\end{itemize}

\item[2.] The laser long term stability has to be improved in order to allow longer integration times. This can be accomplished by actively controlling its operative parameters

\item[3.] improving the scale factor stability and the accuracy with which
it is known, for each ring of the array. This requires:  

\begin{itemize}
\item[a] active control of the geometry of the rings;

\item[b] active control of the relative size of different rings and of their
relative orientation. 
\end{itemize}
\end{itemize}

The main research activities  and tasks  can be grouped in five major areas:

\begin{itemize}
\item[\textit{i)}] The scale factor of each ring must be known and kept
constant at the level of $10^{-10}$. This can be achieved by controlling the
geometry of each ring and the wavelength of the laser, by metrology techniques. "Heterolithic" ring-lasers, i.e. based on a
mechanical design, are cheaper, the mirrors support can be implemented with
suitable translators (usually piezoelectric), and they are flexible enough
to develop complex structures to support rings with different orientations.
We are developing a method which uses information from the ring itself and
the length of its diagonals (which are as well resonant optical cavities),
in order to drive the actuators of the mirrors and keep the ring, from a
geometrical point of view, stable at the required level of $10^{-10}$ \cite{santagata}. To
this aim the heterolithic prototype GP2 has been developed. It
has 6 piezoeletric actuators, and will be the test bench to test the above
mentioned control strategy. We expect to start this experimental work in
2014.

\item[\textit{ii)}] The Lense-Thirring measurement requires to recover the
angular velocity vector, with errors in the relative alignment of the planes
of the rings of the order of 1 nrad. Large frame ring-lasers have been
working with different orientations with respect to the Earth axis, but a
multi-axis system of this size has not been implemented so far. As explained
above the heterolithic ring-laser can "easily" be expanded to hold rings
with different orientation in order to recover the full angular velocity
vector; the very demanding issue is the nrad relative accuracy between
different rings. The octahedron arrangement, in principle a very elegant
design, could be a solution, since the cavities and diagonals can
effectively provide information on the relative angles. The three diagonals
of the octahedron are resonant Fabry-P\'{e}rot linear cavities, that can
monitor the relative angle between two different rings of the octahedron.
Alternative configurations, other than octahedral, and different strategies
for the relative angle monitoring must be investigated, as for example the
use of interferometry with 3D retro-reflectors, which has reached prad level
accuracy \cite{naletto}.

\item[\textit{iii)}] Identify and refine the estimate of the Lamb parameters
which regulate the non-linear dynamics of the ring-laser itself \cite{cuccato}. The
identification procedure will consist of two parts: i) identification and
monitoring of cavity losses from mono-beam amplitudes and phase; ii)
estimate and monitoring of the laser single pass gain and the remaining Lamb
parameters from the measured plasma dispersion function of the He-Ne
mixture. The scale associated to the non linear dynamics permits to perform
the absolute calibration of the instrument. To this aim it is important to
select the most convenient working point of the ring-laser. In addition, the
knowledge of laser dynamics enables us to run a non-linear Kalman filter
which, \textit{a posteriori}, can remove a large fraction of the
backscattering contributions from the rotation rate measurements. In this
way we can improve the long term stability of our rings which represent a
key issue of the GINGER project.

\item[\textit{iv)}] Top quality mirrors are adequate for ring-lasers,
however mirrors will always be a point of concern. The mirrors used for
ring-lasers are standard 1-2 inches substrate, but the quality of the
substrate, the uniformity and quality of the coating are important issues.
Any non-reciprocal effect induced by the mirrors has repercussions in
unbalance in the two counter-propagating beams. So that the mirror
birefringence at the sub ppm level and the possible related problems should
be investigated as well. There are few factories in the world able to
provide this kind of mirrors; we are in touch with all of them.

\item[\textit{v)}] Quantify environmental disturbances in order to have a
proper assessment of the experimental site. This is of paramount importance,
since the ring-laser is an inertial sensor, which is operated to deduce a
global measurement quantity. Therefore the properties of the monument
connecting the ring lasers to the Earth are critical. The operation of the G
ring laser has shown that a near Earth's surface installation is subject to
seasonal changes and noise generated from local wind patterns \cite%
{gebauer2012} \cite{schreiber2013}. Therefore deep underground locations
such as the LNGS have inherent advantages. In general, a deep underground installation within solid rock is almost insensitive to external environmental perturbations. This suggests that LNGS is potentially a good site for installing an array of actively stabilized large ring lasers, but dedicated measurements are necessary.
\end{itemize}

\subsection{Work-plan}
\label{sub-3}

At the time of writing two experimental areas are under construction: GP2,
which will be used to develop the geometry control, has just been
completed, and GINGERino, the 3.6m side ringlaser which will qualify the
LNGS site for GINGER, is under construction. GINGERino will be located in a part of the laboratory away from daily activity, it will be acoustically protected and mounted on top of a granite structure well connected to the ground; it should start taking
data in the second half of 2014.

Further steps will be taken to complete the
characterization of the site, to test the technologies, and to collect
information for geophysics. 

The scientific work-plan toward the GINGER operation can be summarized as
follows:

\begin{itemize}
\item[1)] 2014 - 3.6~m horizontal ring (in principle we should have an
improvement of a factor 7.8 in sensitivity) obtained using the mirror
holders of our first prototype and longer tubes. The size is limited by the room
presently available in the specific location within the LNGS. A different
positioning will allow a larger ring. In any case, as long as the present
mirrors will be used, the side of the ring cannot exceed 6 m,
because the mirrors have a 4m curvature radius. \newline
With GINGERino, the systematics of the laser will be reduced, in particular
backscattering noise should be reduced. In fact, we expect larger biases
from the Earth rotation, the larger distance between mirrors and the gas
discharge, higher Q of the optical cavity (keeping the quality of the mirror
constant). Acoustic shielding and a high-quality reference laser for the
perimeter control are required. The task for GINGERino will be to observe
the Allan deviation of the measurement, in order to understand the
environmental disturbances. We expect to record some relevant seismic
events, and, because of the improvement in sensitivity, also geodetic
signals should be detected. Correlation with G Wettzell measurements should
be possible (common tele-seismic events etc.).

\item[2)] 2015 - one or two smaller rings should be added to GINGERino in
order to reconstruct the angular velocity vector, with $\mu $rad precision
(at least) in the vector direction. In particular, with a ring aligned with the Earth axis a good measurement of the Length of the Day can be pursued.

\item[3)] 2015-2016 the geometry and relative orientations of the rings
should be defined, and it should as well be defined how to monitor the
relative angle between different rings.

\item[4)] 2015-2016 construction of the octahedral arrangement of the full
GINGER experiment

\item[5)] 2017-2018 GINGER in operation.
\end{itemize}

\begin{figure}
\centering
\includegraphics[width=12cm]{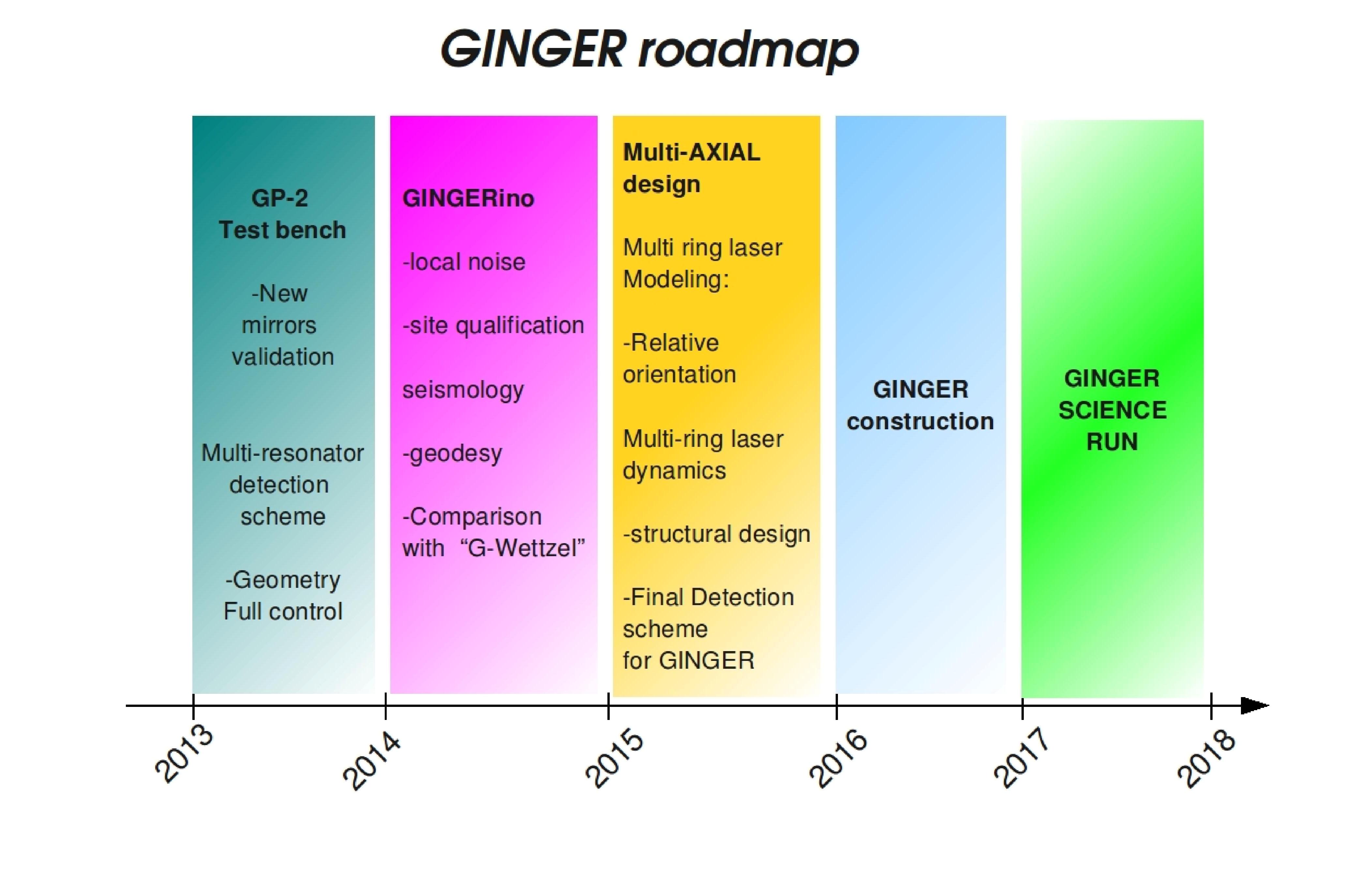}
\caption{Workplan for the GINGER\ roadmap}
\label{fig-2}
\end{figure}

At the end of 2015 it should be possible to qualify the LNGS installation,
and to understand at which level of precision the full installation of
GINGER can be qualified. Fig. ~\ref{fig-2} shows the above outlined
roadmap.

\section{Discussion and Conclusions}

Summing up, we have started a number of practical steps towards the
implementation of an experiment of fundamental physics based on a
three-dimensional array of advanced ring lasers, named GINGER. The first
objective of the experiment is to measure general relativistic effects due
to the rotation of the Earth. In order to overcome the difficulties implicit
in the extreme sensitivity required by the measurement we have build an
international collaboration, involving two more laboratories in the world. A Framework Agreement is being signed between INFN, the University of Canterbury (Christchurch NewZealand), the Technische Universitaet and the Maximilian University of Muenchen.
We have also designed a roadmap (which we have already started to follow)
aimed to test and solve many technological and methodological problems. Step
by step various intermediate facilities are in use and will be built: GP2 to develop the control of the geometry, and GINGERino, based on our first prototype G-Pisa,  to qualify the possible installation inside the underground laboratory of LNGS. In 2015, after the first set of measurements taken inside LNGS, the feasibility of GINGER will be more clear, and its time schedule as well; the construction of the GINGER apparatus $\it{per\ se}$ is rather simple, it should not take more than one or two years. The use of a facility as  LNGS, which is a very large and well equipped laboratory, will facilitate the construction and the start up as well.

We should mention that techniques similar to the ones using lasers could be
envisaged, such as atomic beams interferometry, or long fibers loops \cite{fibres}. Atoms would have the often
stated (intrinsic) advantage of atomic masses much greater than
the photon mass (i.e. much shorter wavelengths), that would allow to reach
the same level of sensitivity for much smaller devices. For the moment this
type of approach has very high potential for atomic 'gyroscopes' but it
cannot (as yet) compete with advanced, large scale ring laser technology.

The terrestrial detection of the Lense-Thirring effect is the main, but not
the only purpose of GINGER. The main difficulty of the Lense-Thirring
measurement is that it corresponds to a constant signal and the calibration
is quite demanding. This is the reason why we are investigating as deeply as
possible the systematics of the laser, and different techniques to extract
the signal: the result could be validated repeating the measurement with
different techniques and operating the laser in a different working point.
The Sagnac effect works as well for a passive cavity, i.e. the measurement
could be repeated with the same apparatus but using an external laser source
to interrogate the array of cavities. The technique of the passive cavity
Sagnac, however, is not mature as the active one. In summary: any
measurement of constant effects is in principle difficult, but a ring-laser
system allows to repeat the measurement with two different methods, which
have different systematics.

Beyond LT we should mention that measurement methods from modern \textit{%
Space Geodesy} perform at about the $10^{-9}$ error level. Lunar Laser
Ranging for example provides precise round trip optical travel times between
a geodetic observatory and cube corner retro-reflectors placed on the moon
by the American APOLLO and the Russian LUNA landers \cite{lunar}. With a long time-series
of observations and continuous technical improvements, which reached a range
precision of several millimeters in recent years, the error margin has
reached a level of $10^{-9}\div 10^{-11}$. As one of many results according
to~\cite{lunar} this led to improved constraints for the gravitational
constant and its spatial and temporal variation of $\dot{G}/G=(2\pm 7)\times
10^{-13}yr^{-1}$ and $\ddot{G}/G=(4\pm 5)\times 10^{-15}yr^{-2}$.

Apart from actually measuring the Lense-Thirring effect with a ground based
gyroscope, also high precision tests of metric theories of gravity in the
framework of the PPN formalism come within reach. With $\hat{J}=I_{\oplus
}\hat{\Omega }_{\oplus }$ according to~\cite{noipr} one obtains
\begin{equation}
\hat{\Omega }_{G}=-(1+\gamma )\frac{GM}{c^{2}R}\Omega _{\oplus }\sin
\vartheta \,\hat{u}_{\vartheta }~,  \label{ppn1}
\end{equation}%
and
\begin{equation}
\hat{\Omega }_{B}=-\frac{1+\gamma +\frac{\alpha _{1}}{4}}{2}\frac{%
GI_{\oplus }}{c^{2}R^{3}}\left[ \hat{\Omega }_{\oplus }-3(\hat{\Omega }%
_{\oplus }\cdot \hat{u}_{r})\hat{u}_{r}\right] ~.
  \label{ppn2}
\end{equation}%
In eq.~\ref{ppn1} and eq.~\ref{ppn2} $\alpha _{1}$ and $\gamma $ represent
the PPN parameters which account for the effect of a preferred reference
frame and the amount of space curvature produced by a unit rest mass. So, high
precision ring laser measurements performed by GINGER should be able to
access $\alpha _{1}$ and $\gamma $. As already stated, being the apparatus on Earth, it should be possible in the future to envisage improvements and upgrading. With improvements of the order of 100-1000, it will be possible to set constraints on the PPN parameters competitive with space experiments.

Georges Sagnac would be surprised to see how far his method has gone after
one century from his initial experiment. His purpose was to prove Special
Relativity wrong, now, under his name, we are preparing the most accurate
verification of one of the effects of General Relativity. Maybe we shall not
prove it wrong but insufficient. We shall know in few years.

\end{document}